\documentstyle[aps,preprint,tighten,floats,epsfig]{revtex}
\begin{document}
\newcommand{\tr}{{\rm tr}}
\newcommand{\sdet}{{\rm sdet}}
\newcommand{\str}{{\rm str}}
\renewcommand{\imath}{i}
\draft
\preprint{hep-th/9606081}
\title{Gauge Dependence in Chern-Simons Theory}
\newcounter{ouraddress}
\author{F.A. Dilkes$^{(\ref{UWO})}%
$\thanks{email: fad@apmaths.uwo.ca} \\
L.C. Martin$^{(\ref{UWO},\ref{Galway})}$%
\thanks{email: lcm@apmaths.uwo.ca} \\
D.G.C. McKeon$^{(\ref{UWO},\ref{Galway})}$%
\thanks{email: tmleafs@apmaths.uwo.ca} \\
T.N. Sherry$^{(\ref{Galway})}$%
\thanks{email: tom.sherry@ucg.ie} \\ 
\vspace{0.5cm}
\refstepcounter{ouraddress}
\label{UWO}
$^{(\ref{UWO})}$Department of Applied Mathematics\\
University of Western Ontario\\
London ~CANADA\\
N6A 5B7\\
\vspace{0.5cm}
\refstepcounter{ouraddress}
\label{Galway}
$^{(\ref{Galway})}$Department of Mathematical Physics\\
University College Galway\\
Galway  IRELAND
\vspace{0.5cm}} 
\date{\verb$Date: 1996/06/14 01:34:19 $}
\maketitle
\begin{abstract}
\noindent
We compute the contribution to the modulus of the one-loop effective 
action in pure non-Abelian
Chern-Simons theory in an arbitrary covariant gauge.  We find that 
the results are dependent on
both the gauge parameter ($\alpha$) and the metric required in the 
gauge fixing.  A contribution
arises that has not been previously encountered; it is of the form 
$(\alpha / \sqrt{p^2}) \epsilon
_{\mu \lambda \nu} p^\lambda$.  This is possible as in three 
dimensions $\alpha$ is
dimensionful.  A variant of proper time regularization is used to 
render these integrals well
behaved (although no divergences occur when the regularization is 
turned off at the end of the
calculation).  Since the original Lagrangian is unaltered in this 
approach, no symmetries of the
classical theory are explicitly broken and $\epsilon_{\mu \lambda 
\nu}$ is handled
unambiguously since the system is three dimensional at all stages of 
the calculation.  
The results are shown to be consistent with
the so-called Nielsen 
identities which predict the explicit gauge parameter dependence using 
an extension of BRS symmetry.
We demonstrate that this $\alpha$ dependence 
may potentially contribute to the vacuum expectation
values of products of Wilson loops.
\end{abstract}

\section{Introduction}

Non-Abelian Chern-Simons theory is an example of a three-dimensional 
topological theory.  It
has been argued that radiative corrections in the theory serve only 
to displace the coupling
constant, despite the fact that the gauge fixing term 
that must be appended to the
classical Lagrangian is metric dependent \cite{Witten}.  
This is consistent 
with a number of (but not all)
perturbative calculations, each of which requires some form of 
regularization \cite{Guadagnini,r3,r4,r5,r6,r7,r8,r9}.  Regulating
this
model is complicated by the three dimensional nature of the tensor
$\epsilon_{\alpha\beta\gamma}$ occurring in the initial Lagrangian, 
and by the fact that the
theory is topological - it is difficult to respect these properties 
if one regulates by altering the
initial Lagrangian in some way.  (For example, addition of a 
Yang-Mills term to the 
Chern-Simons Lagrangian \cite{r3,r5,r8}, while rendering the theory 
super-renormalizable, does not
respect its
topological nature and simple Pauli-Villars regularization 
\cite{Guadagnini}
breaks 
the BRS invariance of the
tree-level effective action.\cite{r6})  A way of circumventing this 
difficulty is to use a variant of operator
regularization
\cite{r10};
in this approach the initial Lagrangian is not modified - rather, the 
operators occurring in the
closed form expression for the generating functional are regulated.

At one loop order, one encounters a functional determinant $\det M$.  
Since $M$ is linear in the
derivative operator, operator regularization cannot be applied 
directly; one must instead consider
${\det}^{\frac{1}{2}}M^2$.  However these two expressions may not be 
equal - a phase
associated with $\det M$ may be lost if $M$ has negative eigenvalues 
\cite{Witten,r11,r12}.  This phase
may
be determined
by considering the so-called $\eta$-function \cite{r13}.  
In \cite{r11} it is 
shown that the contribution of the
$\eta$-function to $\det M$ is 
consistent with the expectations of \cite{Witten}, 
and that this result is
independent of the gauge fixing parameter $\alpha$.  On the other 
hand, in \cite{r14} it has been
demonstrated
that if operator regularization is used to evaluate the contribution 
of the two-point function to
${\det}^{\frac{1}{2}}M^2$ (associated with modulus of $\det M$) 
then a result proportional to
$(g_{\mu\nu} - p_\mu p_\nu / p^2)$ is obtained.  This respects the 
gauge invariance of the
theory, but clearly is not topological in character.  This 
computation was performed in the 
so-called Landau-Honerkamp gauge.

In this paper, we generalize the computation of \cite{r14} by 
including a gauge parameter $\alpha$
(with the Landau gauge recovered in the $\alpha = 0$ limit).  
Operator regularization cannot be used
as in \cite{r14} when $\alpha \neq 0$ due to the form of 
${\det}^{\frac{1}{2}}M^2$; however a simple
modification of the procedure (explained below) allows us to 
circumvent the difficulties which
occur.  It is found that the two-point contribution to 
${\det}^{\frac{1}{2}}M^2$ is in fact dependent
on $\alpha$.  The form of these terms has not been anticipated 
\cite{r8} as the gauge parameter
$\alpha$ is dimensionful; they are proportional to $(\alpha^2 / 
(p^2)^{3/2}) (p^2 g_{\mu \nu}- p_
\mu p _\nu)$ and $(\alpha / \sqrt{p^2})\epsilon _{\mu \lambda \nu} p 
^ \lambda)$.  We now give
the details of the calculation, demonstrate consistency with 
an extension of the BRS 
identities (known as the Nielsen identities), 
and then show that to one-loop order
there may be non-vanishing contributions to the vacuum 
expectation value of a product of Wilson loops that are of 
order~$\alpha$.

\section{The Two-Point Function}
\label{proper2point}

Let us consider the classical Chern-Simons Lagrangian
\begin{equation}
\label{e1}
{\textstyle{\cal {L}}}_{cs} = \frac{1}{2} \epsilon_{\mu\lambda\nu} 
\Big( V_\mu^a
\partial_\lambda
V_\nu^a + \frac{g}{3} f^{abc} V_\mu^a V_\lambda^b V_\nu^c 
\Big) \, .
\end{equation}
If $g=\sqrt{\frac{4\pi}{k}}$ (for some integer 
$k$) then (\ref{e1}) describes a topological quantum field theory 
which is invariant under gauge transformations of arbitrary winding 
number.  We add to it the gauge-fixing Lagrangian 
\begin{equation}
\label{e2}
{\textstyle{\cal {L}}}_g = \frac{-1}{2\alpha} \Big( D_\mu^{am}(A) 
Q_\mu^m \Big)
\Big( D_\nu^{an}(A) Q_\nu^n \Big)
\end{equation}
where $V$ has been split into the sum of classical background field 
$A$ and a quantum field
$Q$\cite{r15}, and $D_\mu^{ab} (A) = \partial_\mu \delta^{ab} + 
gf^{apb} A_\mu^p$.  (Note that in
(2) $\alpha$ is dimensionful.)  The ghost Lagrangian associated with 
(2) is
\begin{equation}
\label{e3}
{\textstyle{\cal{L}}}_{gh} = \overline{c}^a D_\mu^{ab} (A)D_\mu^{bc}
(A+Q)c^c\;\;.
\end{equation}
As this is independent of $\alpha$, its contribution to the one-loop 
generating functional was
computed in \cite{r14} and need not be 
considered further.

Usually the smooth limit to the Landau-Honerkamp gauge requires that 
a first order gauge fixing
formalism be used: i.e. (\ref{e2}) being replaced by
\begin{equation}
\label{e4}
{\textstyle{\cal{L}}}_g = B^aD_\mu^{ab} (A) Q_\mu^b + 
\frac{\alpha}{2}
B^aB^a\;\;.
\end{equation}
The Landau-Honerkamp gauge is obtained now by simply setting $\alpha$ 
to zero (as was done in \cite{Witten}).  However,
when we retain $\alpha$ non-zero throughout the calculations, the 
gauge fixing terms (\ref{e4}) make
the computations prohibitively difficult.  For this reason we choose 
to work with the gauge fixing
Lagrangian (\ref{e2}).

The contribution to the one-loop generating functional coming from 
(\ref{e1}) and (\ref{e2}) is given by
\begin{equation}
\label{e5}
\Gamma^{(1)} = \ln \; {\det}^{-\frac{1}{2}} 
(M_{\mu\nu}^{ab}).
\end{equation}
Unfortunately, the factor
\begin{equation}
\label{e6}
M_{\mu\nu}^{ab} = \epsilon_{\mu\lambda\nu} D_\lambda^{ab}(A) +
\frac{1}{\alpha} D_\mu^{am} (A) D_\nu^{mb}(A)
\end{equation}
is not in a form that is appropriate for applying operator 
regularization \cite{r10,r12} directly
to~(\ref{e5}).  The appropriate procedure in this case \cite{r12} 
is to examine separately the modulus and phase
of $\Gamma^{(1)}$.  The phase calculation, for arbitrary $\alpha$ has 
been computed elsewhere
\cite{r11}.  We now concentrate on the contribution from the modulus, i.e. 
we examine
\begin{mathletters}
\begin{eqnarray}
\mid \Gamma^{(1)} \mid\;\; & = & \;\; \ln {\det}^{-
\frac{1}{4}}\Big(M_{\mu\kappa}^{ap}
M_{\kappa\nu}^{pb} \Big) \label{e7} \\
& = & -\frac{1}{4} \tr \ln \Big(M_{\mu\kappa}^{ap} 
M_{\kappa\nu}^{pb}
\Big) \label{e8} \\
& = & \frac{1}{4} \int_0^\infty 
\frac{dt}{t} \tr\;\Big(
e^{-M^{2}t}\Big) \;\;.\label{e9}
\end{eqnarray}
\end{mathletters}%

A perturbative expansion of $\mid \Gamma^{(1)} \mid$ can be developed 
if we first partition
$M^2$ into the sum $M_0^2 + M_1^2$ where $M_1^2$ contains all the 
dependence on the
background field $A_\mu^a$, and secondly use the Schwinger expansion 
\cite{r10,r16}
\begin{eqnarray}
\tr \, e^{-M^2t} &
= & \tr \Big[e^{-M_{0}^{2}t} 
+ (-t) e^{-M_{0}^{2}t} M_1^2 \label{e10} \\
& &
+ \frac{(-t)^2}{2} \int_0^1 du \; e^{-(1-u)M_{0}^{2}t} M_1^2 e^{-u 
M_{0}^{2}t} M_1^2
+ ... \Big] \nonumber
\end{eqnarray}
As we are interested in the two point function 
we need only retain terms which are 
second order in the background field $A_\mu^a$.

From (\ref{e6}) we see that
\begin{equation}
\label{e11}
(M_0^2)_{\mu\nu}^{ab} = \delta^{ab} \Big[ p^2 \delta_{\mu\nu} - 
p_\mu
p_\nu +
\frac{1}{\alpha^2} p^2 p_\mu p_\nu \Big] \;\; , \; (p \equiv -i 
\partial)
\end{equation}
so that
\begin{equation}
\label{e12}
e^{-M_{0}^{2}t} = e^{-p^{2}t} T_{\mu\nu} +
e^{-\frac{p^{4}t}{\alpha^2}}
L_{\mu\nu}
\end{equation}
{\centering where $T_{\mu\nu} \equiv \delta_{\mu\nu} - p_\mu p_\nu / 
p^2$, and $L_{\mu\nu} \equiv
p_\mu p_\nu /p^2$.\\}

\noindent (In deriving (\ref{e12}), we have used that fact that $T$ and $L$ 
constitute a complete set of orthogonal projection operators.)

The presence of the factor $p^4$ in the second exponential in 
(\ref{e12}) makes implementation of
operator regularization as outlined in \cite{r10} prohibitively 
difficult.  We can however make use of the formula
\begin{eqnarray}
& &\frac{1}{a^mb^n} 
\int_0^\infty dt\;\;t \int_0^1 du \;\;e^
{-[\frac{a^{p}}{A} (1-u) + \frac{b^{q}}B\;u]t} \label{e13} \\ 
& &= \frac{AB}{\Gamma(m+p)\Gamma(n+q)} \int_0^1 dx\;\; x^{m+p-1} (1-
x)^{n+q-1}
\int_0^\infty d\tau \;\; \tau^{m+n+p+q-1}\;e^{-[ax+b(1-x)]\tau} 
\nonumber \end{eqnarray}
in order to eliminate dependence on $p^4$ that may occur in the 
arguments of exponentials
appearing in (\ref{e10}).  In order to regulate any potential 
divergences in integrals appearing at this stage in the calculation,
the factor of $\tau^{m+n+p+q-1}$ in (\ref{e10}) is replaced by
$\tau^{\lambda+m+n+p+q-1}$; the regulating parameter $\lambda$ is 
allowed to pass to the limit zero at the end of the calculation.  
In the present situation, no divergence arises in the final
expression in this $\lambda \rightarrow 0$ limit, so in principle the 
parameter $\lambda$ is
superfluous, although it does serve to render all intermediate 
expressions in the calculation well defined.

In (\ref{e13}) we take $p+m$ and $q+n$ to be greater than zero, 
otherwise we would just be left with tadpole integrals of the form
$$\int d^3p \frac{1}{(p^2)^a} \;\;\;\; , \;\;(a > 0)$$
which are taken to be zero \cite{r14}.

We now outline the steps used to obtain the contribution from $\mid 
\Gamma^{(1)} \mid$ to the
two-point function to ${\det}^{-\frac{1}{4}} M^2$.  First, upon making
use of the expansion (\ref{e10})
in (\ref{e9}) with operator $M$
defined in (\ref{e6}) we obtain
$$\mid \Gamma_{AA}^{(1)} \mid \;\;=\;\; g^2 C_2 \int_0^\infty dt\;t 
\int_0^1  du \int
\frac{d^3p \; d^3q}{(2\pi)^3} \Bigg\{e^{-[p^2(1-u)+q^2u]t}$$
$$\Bigg[ -\frac{(q \cdot p)^2}{q^2} A(+) \cdot A(-) + \Big( q \cdot
A(+) p \cdot A(-) + q \cdot A(-) p \cdot A(+) \Big) \big( 1 + \frac{q
\cdot p}{q^2} \big) \Bigg]$$
$$+ e^{-[q^2(1-u) + \frac{p^4}{\alpha^2} u]t} \Bigg[ A(+) \cdot A(-)
\frac{(q \cdot p)^2}{p^2} + q \cdot A(+) q \cdot A(-)$$
\begin{equation}
\label{e14}
-\frac{q \cdot p}{p^2} \Big(q \cdot A(+) p \cdot A(-) + q \cdot
A(-)p \cdot A(+) \Big)
\end{equation}
$$+ \frac{i}{\alpha} \Big( q \cdot A(-) q \cdot p \times A(+) - q
\cdot A(+) q\cdot p \times A(-) + 2q \cdot p (q-p) \cdot A(+)
\times A(-) \Big)$$
$$+ \frac{p^2}{\alpha^2} \Big( -4 q \cdot p A(+) \cdot A(-) + 2q \cdot
A(+) p \cdot A(-) + 2q \cdot A(-) p \cdot A(+)$$
$$\;\;\;\;\;\;\;\;\;\;\;\;\;\;-p \cdot A(-) p \cdot A(+) - q \cdot A(-
) q\cdot A(+) -
\frac{1}{q^2} q \cdot p \times A(+) q \cdot p \times A(-) \Big)
\Bigg]$$
$$+ e^{-[q^4(1-u) + p^4u]t/ \alpha^2} \Bigg[\frac{1}{\alpha^2} \big(1 
+ \frac{p^2}{q^2} \big)
q
\cdot p \times  A(+) q \cdot p \times A(-)$$
$$ + \frac{i}{\alpha^3} q \cdot p \times A(+) \Big( q^2p \cdot A(-) + 
p^2 q \cdot A(-) + (p^2
+ q^2)(q+p) \cdot A(-) \Big)$$
$$+\frac{1}{\alpha^4} p^2q^2 \Big(p \cdot A(-) p \cdot A(+) + q \cdot 
A(+) p \cdot A(-) + q
\cdot A(-) p\cdot A(+) \Big) \Bigg] \Bigg\}\;.$$
In arriving at (\ref{e14}), we have computed functional traces in 
momentum space \cite{r10,r16} so that 
\begin{equation}
\label{e15}
(2 \pi)^{3/2} \langle p \mid A_\mu^a \mid q \rangle = A_\mu^a (p-
q)\;\; .
\end{equation}
We have used the notation $f^{amn} f^{bmn} = C_2 \delta^{ab}$, 
$A_\mu(\pm) \equiv A_\mu^a
(\pm (q-p))$ and $X \cdot Y \times Z \equiv 
\epsilon_{\alpha\beta\gamma} X_\alpha Y_\beta
Z_\gamma$.  All contributions to the two-point function which are 
eventually proportional to
tadpole integrals have been dropped in (\ref{e14}).

By using (\ref{e13}), (\ref{e14}) can be converted 
into a sum of integrals in which 
the argument of the
exponential occurring in each integrand is just $[xp^2 + (1-
x)q^2]\tau$, allowing us to use the
standard integrals to compute $\Gamma_{AA}^{(1)}$, as was done in the 
examples given in
\cite{r10}. 
Some tedious calculation leaves us with the result
\begin{equation}
\label{e16}
|\Gamma_{AA}^{(1)}| = g^2\frac{C_2}{32} \int d^3p \sqrt{p^2} \Bigg\{  
\Big(g_{\mu\nu}
-
\frac{p_\mu p_\nu}{p^2}\Big) \Big(1 + \frac{\alpha^2}{4p^2}
\Big) + \frac{i \alpha}{p^2}
\epsilon_{\mu\lambda\nu} p_\lambda \Bigg\} A_\mu^a (p) A_\nu^a (-
p)\;.
\end{equation}
This is transverse, as is required by gauge invariance, but 
is manifestly dependent on both
the metric and gauge parameter $\alpha$.  Neither of the terms 
dependent on $\alpha$ have been
anticipated \cite{r8}.  
The dependence on $\alpha$ occurs in the transverse direction in 
(\ref{e16}), which is unlike what occurs in Yang-Mills theory where
gauge-parameter dependence is confined to the 
unphysical longitudinal sector
of the vector field.  This 
does not imply that the propagation of physical degrees
of freedom depends on an arbitrary parameter, as the pure Chern-Simons
theory in fact has
no dynamical content!

\section{Nielsen Identities}
\label{Nielsenidentities}

We now turn to the Nielsen identities to test the consistency of 
our expression for 
$\Gamma^{(1)}_{AA}$.  Much as the Ward-Takahashi and 
Slavnov-Taylor identities
arise from considering the BRS symmetry of the Lagrangian, one 
obtains the
Nielsen Identities \cite{r17,Kluberg} by extending this symmetry to 
include 
variations in the gauge parameter.  In doing so we are able to obtain relations between the
variation of the gauge parameter in a given Green's function with 
their product of other Green's functions (e.g. $\frac{\partial 
}{\partial \alpha} \Gamma_1=
\Gamma_2 \Gamma_3$).  The relevant Nielsen identity for 
$\Gamma^{(1)}_{AA}$ will
be derived and the other Green's functions arising in the identity 
will be
calculated.  We will then show that the $\alpha$-dependence 
in (\ref{e16})
is consistent with this  
identity. 

\begin{figure}
\centering
\epsfig{file=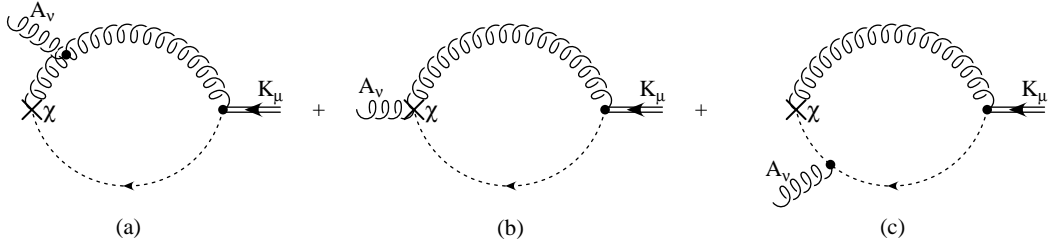}
\centering
\caption{The one-loop contributions to $\Gamma_{\chi K_\mu A_\nu}$.  
(Curly and dashed lines indicate gauge bosons and ghosts respectively.) 
\label{figoneloop}}
\end{figure}

In the background field formalism the full Lagrangian for a 
non-Abelian 
gauge field (including all sources) \cite{Kluberg} is given by 

\begin{eqnarray}
\label{fullLagrangian}
{\cal {L}} & = & {\cal {L}}_{cs} (A+Q)-
\frac{1}{2\alpha} \left( D_\mu(A) Q_\mu \right) 
\left( D_\nu(A) Q_\nu \right)+
\overline{c} D_\mu (A)D_\mu (A+Q)c
                          \hspace{10mm} \nonumber \\
&& \hspace{10mm}
+ K_\mu D_\mu\left(A+Q\right)c
+ \frac{g}{2} K \cdot c \times c
+ L_\mu \left(D_\mu (A+Q )\overline{c} - K_\mu \right) \, ,
\end{eqnarray}
where $K \cdot c \times c = K^a f^{abc} c^b c^c$ and
 other group indices have been suppressed.
The Lagrangian is invariant under the usual BRS transformations:

\begin{eqnarray}
\delta A_\mu  =  -L_\mu \lambda &  \hspace{20mm} & 
\delta Q_\mu  =  D_\mu (A+Q) c \lambda + L_\mu \lambda  \nonumber \\
\delta c = \frac{g}{2}  c \times  c \lambda & \hspace{20mm} &
\delta \overline{c} = \frac{-1}{\alpha} D_\mu (A) Q_\mu \lambda 
\end{eqnarray}
where $\lambda$ is a global Grassmann parameter.  By adding the 
following term parametrized by the global Grassmann variable $\chi$ 
(which does not alter the dynamics) to the Lagrangian 

\begin{equation}
\label{chi}
{\cal L} \rightarrow {\cal L} + \frac{\chi}{2\alpha}\overline{c} 
D_\mu(A) Q_\mu 
\, ,
\end{equation}
the new Lagrangian is invariant under the following extended set of 
BRS transformations~\cite{r17,Kluberg}
\begin{eqnarray}
\label{variations}
\delta A_\mu  =  -L_\mu \lambda &  \hspace{20mm} & 
\delta Q_\mu  =  D_\mu (A+Q) c \lambda + L_\mu \lambda  \nonumber \\
\delta c = \frac{g}{2}  c \times c \lambda  
& \hspace{20mm} &
\delta \overline{c} = \frac {-1}{\alpha} D_\mu (A) Q_\mu \lambda + 
   \frac {\chi}{2\alpha} \overline {c} \lambda \\
\delta \alpha = \chi \lambda & \hspace{20mm} & 
\delta \chi = 0 \, .\nonumber 
\end{eqnarray}
The generating functional is 
\begin{equation}
Z = \int [{\cal D}  Q_\mu  \, {\cal D}\overline{c} \, {\cal D}c] \, 
e ^ {  -\int d^3x \left( {\cal L} + 
J_\mu Q_\mu + \overline{J}_c c + \overline{c} J_{\overline c} \right) } \, ,
\end{equation}
and the 1PI generating functional is 
\begin{eqnarray}
\Gamma [Q_\mu,c,\overline{c},A_\mu,L_\mu,K_\mu,K,\alpha,\chi] 
& \equiv &
W[J_\mu,\overline{J}_c,J_{\overline c},A_\mu,L_\mu,K_\mu,K,\alpha,
\chi]
                                                     \nonumber \\
& & -\int d^3x \, 
     \left[ J_\mu Q_\mu + \overline{J}_cc + 
           \overline{c} J_{\overline{c}} \right]
\end{eqnarray}
where $ W \equiv - \ln Z$.  $\Gamma$ is invariant in the
following manner
\begin{eqnarray}
\label{invariantgamma}
\delta \Gamma & \equiv
              & \delta Q_\mu \frac{\delta \Gamma}{\delta Q_\mu}
  + \delta c  \frac{\delta \Gamma}{\delta c}
  + \delta \overline{c} \frac{\delta \Gamma}{\delta \overline{c}}
  + \delta  A_\mu \frac{\delta \Gamma}{\delta A_\mu}
  + \delta \alpha  \frac{\delta \Gamma}{\delta \alpha} \\
& = & 0 \nonumber \, . 
\end{eqnarray}
Note that integration over coordinate space is understood. 
The quantum fields here are not the ones which originally appear in
the Lagrangian but rather the Legendre transform fields (i.e. 
vacuum expectation 
values of the Lagrangian fields) which appear 
in the proper generating  functional.  Thus the BRS variations in 
terms of the these fields are not necessarily 
given by (\ref{variations}) but may be 
expressed in terms of derivatives with respect to the sources 
(this can be shown to hold for composite fields as well).
Also, for the particular Green's functions we will want to obtain, we 
may choose to set $L_\mu = 0$
(i.e. $\delta A_\mu =0$).  Thus equation (\ref{invariantgamma})
becomes 
\begin{equation} 
0 = \frac{\delta \Gamma}{\delta K_\mu} 
    \frac{\delta \Gamma}{\delta Q_\mu}
  - \frac{\delta \Gamma}{\delta K}
    \frac{\delta \Gamma}{\delta c}
  + \left[ 
       \frac{1}{\alpha} D_\mu(A) Q_\mu 
       - \frac{\chi}{2\alpha} \overline{c} 
    \right]
    \frac{\delta \Gamma}{\delta \overline{c}}
  + \chi \frac{\delta \Gamma}{\delta \alpha}
\, ;
\end{equation}
differentiating with respect to $\chi$ and then setting $\chi$ equal 
to $0$
we obtain
\begin{equation}
0 = \frac{\delta^2 \Gamma}{\delta \chi \delta K_\mu}
    \frac{\delta \Gamma}{\delta Q_\mu}
   -\frac{\delta \Gamma}{\delta K_\mu}
    \frac{\delta^2 \Gamma}{\delta \chi \delta Q_\mu}
   -\frac{\delta^2 \Gamma}{\delta \chi \delta K}
    \frac{\delta \Gamma}{\delta c} 
   -\frac{\delta \Gamma}{\delta K}
    \frac{\delta^2 \Gamma}{\delta \chi \delta c}  
   +\frac{1}{\alpha} D_\mu(A) Q_\mu
    \frac{\delta^2 \Gamma}{\delta \chi \delta \overline{c}}
   -\frac{1}{2 \alpha} \overline{c} \frac{\delta \Gamma}{\delta \overline{c}}
   +\frac{\delta \Gamma}{\delta \alpha}
\, .
\end{equation}
Since we take no further derivatives with respect to the 
ghost fields or sources $K$ we may set them to zero.  
Ghost number conservation then implies 
that all but the following terms vanish,
\begin{equation}
0 = \frac{\delta^2 \Gamma}{\delta \chi \delta K_\mu}
    \frac{\delta \Gamma}{\delta Q_\mu}
   +\frac{1}{\alpha} D_\mu(A)
    Q_{\mu}
    \frac{\delta^2 \Gamma}{\delta \chi \delta \overline{c}}
   +\frac{\delta \Gamma}{\delta \alpha} \; .
\end{equation}
We may now
set $Q_\mu = 0$ and differentiate the two remaining terms with 
respect to  
$A_\lambda(x)$ and $A_\nu(y)$ to obtain
\begin{eqnarray}
0 & = &
    \frac{\delta^4 \Gamma}
   {\delta \chi \delta K_\mu  \delta A_\nu(y) \delta A_\lambda(x) }
    \frac{\delta \Gamma}{\delta Q_\mu }
+ \frac{\delta^3 \Gamma}{\delta \chi \delta K_\mu \delta A_\lambda(x)}
    \frac{\delta^2 \Gamma}{\delta A_\nu(y) \delta Q_\mu}
 +  \frac{\delta^3 \Gamma}{\delta \chi \delta K_\mu \delta A_\nu(y) }
    \frac{\delta^2 \Gamma}{\delta A_\lambda(x) \delta Q_\mu }
 \nonumber \\
& &  +  \frac{\delta^2 \Gamma}{\delta \chi \delta K_\mu }
  \frac{\delta^3 \Gamma}{\delta A_\nu(y) \delta A_\lambda(x) 
  \delta Q_\mu }
 +  \frac{\delta^3 \Gamma}{\delta A_\nu(y) \delta A_\lambda(x) 
   \delta \alpha }
  \, .
\end{eqnarray}
There is a contribution from neither 
$\frac{\delta \Gamma}{\delta Q_\mu^a}$
 nor from $\frac{\delta^2 \Gamma}{\delta \chi \delta K_\mu^a}$
 as these terms have a 
single uncontracted group index and $\Gamma$ is invariant 
in group space. 
Thus we are left with the {\em Nielsen identity}
\begin{eqnarray}
\label{Nielsenid}
0 = 
 \frac{\delta^3 \Gamma}{\delta \chi \delta K_\mu  \delta A_\lambda(x)}
    \frac{\delta^2 \Gamma}{\delta A_\nu(y) \delta Q_\mu}
 +  \frac{\delta^3 \Gamma}{\delta \chi \delta K_\mu \delta A_\nu(y) }
    \frac{\delta^2 \Gamma}{\delta A_\lambda(x) \delta Q_\mu }
 +  \frac{\delta^3 \Gamma}{\delta A_\nu(y) \delta A_\lambda(x) \delta
 \alpha }
  \, .
\end{eqnarray}
This equation is trivially satisfied at tree-level, since the proper
vertex $\frac{\delta^3\Gamma}{\delta\chi \delta K_\mu 
\delta A_\lambda}$
have no tree-order contribution as is evident 
from~(\ref{fullLagrangian}).
In fact, the identity~(\ref{Nielsenid}) can be checked at one-loop
order without having to calculate radiative corrections to 
$\frac{\delta^2 \Gamma}{\delta Q \delta A}$.  We thus turn our
attention to the calculation of one-loop contributions to 
$\frac{\delta^3 \Gamma}{\delta \chi \delta K_\mu \delta A_\nu}$, pictured in
figure~\ref{figoneloop}.
Power counting arguments show 
that these loop integrals are not divergent and so no 
regulating procedure
will be required.  The relevant terms in the extended Lagrangian 
bilinear in the quantum fields can be written as follows;
\begin{eqnarray}
{\cal L}^{(2)} = \frac{1}{2}
\left( Q_\mu^a \overline{c}^a c^a \right) & 
{\cal M}_{\mu,\nu}^{ab} 
&
\left(
\begin{array}{lll}
Q_\nu^b  \\
\overline{c}^b \\
c^b
\end{array}
\right)
\end{eqnarray}
with
\begin{eqnarray}
{\cal M}_{\mu,\nu}^{ab} =
&
\left[
\begin{array}{ccc}
\epsilon_{\mu \lambda \nu} D_\lambda^{ab}(A) +
\frac{1}{\alpha} D_\mu^{am}(A) D_\nu^{mb}(A) &
\frac{-1}{2 \alpha} \chi D_\mu^{ab}(A) &
 f^{abc} K_\mu^c \\
\frac{-1}{2 \alpha}\chi 
D_\nu^{ab}(A) & 0 & D_\lambda^{am}(A) D_\lambda^{mb}(A) \\
f^{abc} K_\nu^{c} & -D_\lambda^{am}(A) D_\lambda^{mb}(A) & 0  
\end{array} 
\right] \, .
\end{eqnarray}
Then the one loop effective action for $K_\mu$, $\chi$ and $A_\mu$ is
given by
\begin{mathletters}
\begin{eqnarray}
\Gamma^{(1)} & = & \ln {\sdet}^{- \frac{1}{2}} {\cal M}  \\ 
             & = & - \frac{1}{2} \str \ln {\cal M} \, .
\end{eqnarray}
\end{mathletters}%
We split up ${\cal M}$ in the following manner
\begin{equation}
\label{lnM}
\ln{\cal M} = \ln{\left({\cal M}_O + {\cal M}_I\right)}
\end{equation}
where
\begin{eqnarray} {\cal M}_O &  = & 
\left[
\begin{array}{ccc}
\epsilon_{\mu \lambda \nu} D_\lambda^{ab}(A) + 
\frac{1}{\alpha} D_\mu^{am}(A) D_\nu^{mb}(A) & 0 & 0   \\
0 & 0  & D_\lambda^{am}(A) D_\lambda^{mb}(A)  \\
0 & -D_\lambda(A)^{am} D_\lambda^{mb}(A) & 0 
\end{array} 
\right] 
\end{eqnarray}
and
\begin{eqnarray} {\cal M}_I & = & 
\left[
\begin{array}{ccc}
0 & \frac{-1}{2 \alpha} \chi D_\mu^{ab}(A) & f^{abc} K_\mu^c \\
\frac{-1}{2 \alpha} \chi D_\nu^{ab}(A) & 0  & 0  \\
f^{abc} K_\nu^{c} &  0 & 0  
\end{array} 
\right] \, .
\end{eqnarray}
Writing (\ref{lnM}) as
\begin{equation}
\ln  ({\cal M}_O (1 + {\cal M}_O^{-1} {\cal M}_I)) = 
  \ln {\cal M}_O + \ln(1+ {\cal M}_O ^{-1} {\cal M}_I) \, ,
\end{equation}
we see that the first term contains only the 
background field $A$ and thus will not 
contribute to diagrams with external $\chi$'s and $K_\mu$'s.
Expanding the second term in a series, 
one finds that only the quadratic term
contains factors of $\chi K_\mu A_\nu$ 
and thus contributes to the relevant three-point 
functions.  Thus
\begin{equation}
\Gamma^{(1)}_{\chi K_\mu}[A] = 
  \frac{1}{4} \str ({\cal M}_0^{-1} {\cal M}_I)^2 \,.
\end{equation}
Finding the required inverses and
taking the supertrace we get that in
momentum space $\Gamma^{(1)}_{\chi{K_\mu}{A_\nu}}$ is given by those
terms in the expression 
\begin{eqnarray}
& &
  \tr \frac{\chi}{2 \alpha}
  \int \frac {d^3p \, d^3q} {\left(2 \pi \right)^3}  
  \left[
  \left( \imath \epsilon_{\kappa \beta \lambda} 
  \frac {p_\beta }{p^2} - \alpha \frac 
{p_\kappa p_\lambda}{p^4} \right)
  -
  g \left( \imath \epsilon_{\kappa \beta \sigma} 
  \frac {p_\beta }{p^2} - \alpha \frac 
{p_\kappa p_\sigma}{p^4} \right)
  \right. \nonumber \\
 & & \left. \times
  \left(\frac {\imath}{\alpha}  
  \left(p_\sigma A_\pi(-) + A_\sigma(-) p_\pi \right)
  + \epsilon_{\sigma \omega \pi} A_\omega(-) \right)  
  \left( \imath \epsilon_{\pi \delta \lambda } 
  \frac {p_\delta }{p^2} - \alpha \frac {p_\pi p_\lambda}{p^4} \right)
  \right]  \nonumber \\ 
  & & 
  \biggl[ \imath p_\lambda + g A_\lambda(-)  \biggr] 
  \left[ \frac{-1}{p^2} - \frac{1}{p^2} \imath g 
  \left( p_\mu A_\mu(-) + A_\mu(-) p_\mu \right) \frac{1}{p^2} \right]
   g  K_\kappa(+) \label{Nielsenoneloop}  
\end{eqnarray}
which are linear in $A_\nu$.

One can readily see that upon expanding, 
the term with the fields $A$ coming
from the first, second and third set of 
square brackets correspond to the
Feynman diagrams shown in Fig. \ref{figoneloop} 
(a), (b), and (c) respectively.
The remaining two point functions 
appearing in equation (\ref{Nielsenid})
are simply

\begin{equation}
\label{AQtree}
\frac{\delta^2 \Gamma^{(0)}}{\delta A_\mu(x) \delta Q_\nu(y)} = 
         \epsilon_{\mu \lambda \nu} \partial_\lambda^y 
         \delta (y-x) \, .
\end{equation}
Upon using the expressions for $\Gamma_{AA}^{(1)}$, 
$\Gamma_{\chi K_\mu A}^{(1)}$ and $\Gamma_{A Q}^{(0)}$ 
from equations (\ref{e16}), (\ref{Nielsenoneloop}) and (\ref{AQtree})
respectively, 
we see that the Nielsen identity 
of (\ref{Nielsenid}) is indeed satisfied to order 
$g^2$.

\section{Topological Observables}
\label{topologicalobservables}

In the section~\ref{proper2point} we derived the existence of
gauge-parameter dependence in the 
two-point correlation $\langle V(x) V(y) 
\rangle$.  Although these terms do not seem to have appeared before
in the literature, we should not be too surprised by their existence
since, as shown section~\ref{Nielsenidentities}, this dependence is  
just a consequence of the BRS variation of operator
$V(x)V(y)$.

We now turn our attention to the 
matter of gauge-parameter dependence in the topological
invariants of the theory.
Topological observables have been of primary
interest in the development of Chern-Simons theory 
(eg. see \cite{Witten,Guadagnini,Polychronakos}).  
It is well-understood that
the conventional gauge-fixing procedure of Faddeev and Popov requires
the introduction of a metric in the field dynamics and breaks,
not only the local gauge invariance, but also the topological
invariance of the theory.  However, as argued 
in \cite{Birmingham}, the       
topological invariance can be shown to be recovered on the physical 
(BRS invariant) space.  
We show in this section that the interplay
between gauge-fixing and topology may play another role.

The usual observables in Chern-Simons theory are the non-local 
metric-independent {\em Wilson
link} operators whose expectation values are given by
\begin{equation}
\label{Wilsonlink}
\label{e17} 
\langle W[L]\rangle = 
\left\langle 0 
\left| T \prod_{C_R \in L} W[C_R] \right| 0 \right\rangle \,\,\, ,
\end{equation}
where $L$ can be any non-intersecting set of knots $\{C_R\}$, and
\begin{mathletters}
\label{Wilsonloop}
\begin{eqnarray}
W[C_R] & = & 
\tr_R {\cal P} \exp \left\{ ig \oint_C dx^\mu \cdot V_\mu(x) 
\right\} \\
\label{Wilsonloopexpansion}
& \equiv & \tr_R \sum_{n=0}^\infty (ig)^n \oint_C dx_n \cdot V(x_n) 
\int_x^{x_n} dx_{n-1} \cdot V(x_{n-1}) \cdots 
\int_x^{x_2} dx_1 \cdot V(x_1) 
\end{eqnarray}
\end{mathletters}%
is the {\em Wilson loop} operator  
of an 
oriented closed curve 
$C$ (starting at an arbitrarily chosen point $x$) 
and associated Lie algebra representation $R$;
${\cal P}$ refers to ``path ordering''.

The fundamental problem which occurs
in trying to identify the Wilson loop
operators with topological 
gauge-independent observables arises from the 
regions of the integrals in (\ref{Wilsonloop}) where
two or more of the gauge fields occur at a coincident point.  
It is well
known in quantum field theory that composite quantum fields
require special regularization over 
and above any regularization of the 
elementary fields (see for example Collins \cite{Collins}).  Although,
these coincident contributions do not 
introduce any further divergences
into the expectation value of $W[C]$ 
(owing to the zero-measure of the
region over which they contribute) they do serve to destroy the 
topological properties which are enjoyed by $W[C]$ 
at the classical level.
It is thus found that, while the expectation 
values of cross-terms between the
various elements of the 
link in (\ref{Wilsonlink}) have an {\it a priori}
 well-defined topological meaning, the 
{\em self-interacting} terms 
(those involving more than one power on the 
gauge-connection on any given knot) do not.

\begin{figure}[htb]
\centering
\psfig{file=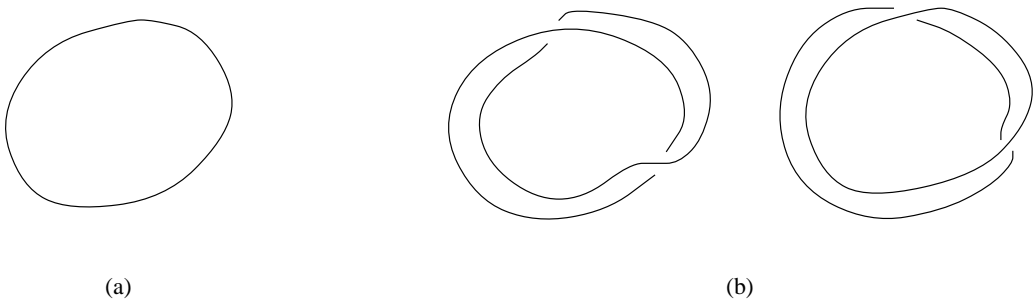}
\centering
\caption{The unknot (a) is shown here together with two topologically 
distinct framings (b).
\label{ribbon}}
\end{figure}

The solution which is typically proposed is as follows.  Instead of
thinking of a given knot $C_R$ as a 1-manifold, 
we can allow them
to have a slightly extended structure: that of a ribbon, 
$\tilde C_R$ as shown in 
figure~\ref{ribbon}.  This construction (referred to as {\em framing})
has been used extensively \cite{Polychronakos,Guadagnini}
to give topological meaning 
to the Wilson line operator which
will thus depend on the topological structure of the ribbon.  

\begin{figure}
\centering
\psfig{file=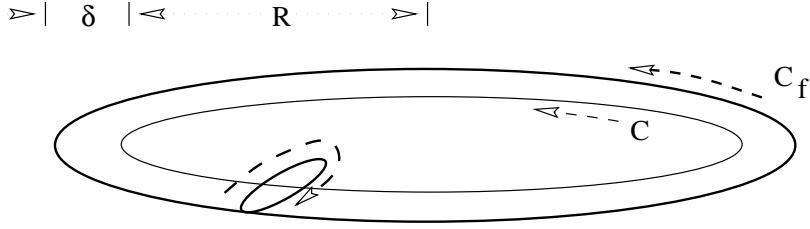}
\centering
\caption{Linked framing for a circular unknot.
\label{exampleframe}}
\end{figure}

At this point we turn our 
attention to the leading $\alpha$ dependence of
the two-point contribution to the expectation value   
(\ref{Wilsonlink}) resulting from 
tree-level effects and the radiative 
correction given by (\ref{e16}).  
For this we shall content ourselves with a trivial link $L$ composed, 
for example, of 
only a single circular unknot $\{ C_R \}$.
Since, for the moment, 
we are concerned only with the $n=2$ term of 
(\ref{Wilsonloopexpansion}) 
it will be sufficient to make a simple implementation of the framing
procedure given by
\begin{equation}
\label{circleframe}
\label{e20}
T^{(2)}[C,C_f]_R = \tr_R \case{(ig)^2}{2!} \oint_C dx^\mu  
\oint_{C_f} dy^\nu \langle 0 | V_\mu(x) V_\nu(y) | 0 \rangle 
\,\, .
\end{equation}
where the framing $\tilde C$ has been reduced to two curves $C$ (the
original knot) and $C_f$ (see, for instance, 
figure~\ref{exampleframe}).  $T^{(2)}$ refers to the contribution to
$\langle W[L] \rangle$ in (\ref{Wilsonlink}) coming from a single Wilson
loop $C$ with framing $C_f$ that is of second order in the vector field.

Now, the propagator for Chern-Simons theory is, 
from (\ref{e1}) and (\ref{e2}),
proportional to
\begin{equation}
\label{e18}
\Lambda_{\mu\nu}^{ab} = 
\left[ \frac{i\epsilon_{\mu\lambda\nu}p_\lambda}{p^2}
-\alpha\frac{p_\mu p_\nu}{p^4} \right] \delta^{ab}\, .
\end{equation}
Let us first consider the contribution of the term in (\ref{e18}) of
order $\alpha$ to (\ref{e20}).   
The tree-level two-point function in coordinate space is 
just the Fourier transform of (\ref{e18}).  From invariance arguments
we must have
\begin{equation}
\label{e21}
\int d^3p \, \frac{p_\alpha p_\beta}{p^4} 
e^{i\overline{p} \cdot \overline{r}} = A
\frac{\delta_{\alpha\beta}}{r} + B \frac{r_\alpha r_\beta}{r^3} \, ,
\end{equation}
and thus by contracting (\ref{e21}) with $\delta_{\alpha\beta}$ 
and $r_\alpha r_\beta$ we obtain
\begin{mathletters}
\label{Fouriercomp}
\begin{equation}
\label{e22}
\frac{3A+B}{r} = \frac{4\pi}{r} \int^\infty_0 dx \frac{\sin x}{x} 
= \frac{2\pi^2}{r} \, ,
\end{equation}
\begin{eqnarray}
(A+B)r & = & 4\pi r \int^\infty_0 dx \left[ \frac{\sin x}{x} 
+ 2 \left( \frac{\cos x}{x^2} - \frac{\sin x}{x^3} \right) \right] 
\nonumber \\
& = & 0 \, , \label{e23}
\end{eqnarray}
\end{mathletters}%
so that $A=-B=\pi^2$.  Hence, by (\ref{e18} - \ref{Fouriercomp}),
to order $\alpha$, 
and to zero-loop order
in the loop expansion, (\ref{e20}) is proportional to 
the integral $I^{(0)}$ given by
\begin{eqnarray}
I^{(0)} & 
= & 
\pi^2 \oint_C dx_\lambda \oint_{C_f} dy_\sigma 
\left[ \frac{\delta_{\lambda\sigma}}{|\overline{x}-\overline{y}|} - 
\frac{(x-y)_\lambda (x-y)_\sigma}{|\overline{x}-\overline{y}|^3} 
\right]
                       \nonumber \\
& = & \pi^2 \oint_C dx_\lambda \oint_{C_f} dy_\sigma 
\frac{\partial^2}{\partial x^\lambda \partial x^\sigma} |\overline{x}-
                                   \overline{y}| = 0 \, .
\label{e24}
\end{eqnarray}
Since, in (\ref{e24}) we have an integral around a closed loop of a 
gradient, it vanishes.  Thus, to the order we have considered,
there is no $\alpha$ dependence in $\langle W[C] \rangle$, as defined
above.
We now consider the effect of including the result of (\ref{e16}) on 
$T^{(2)}$ in (\ref{e20}).  

To order $\alpha$, the one-loop contribution to 
(\ref{e20}) involves having
to consider, by (\ref{e16}) and (\ref{e18}),
\begin{equation}
\label{e25}
\left( \frac{i\epsilon_{\mu\lambda\alpha}p_\lambda}{p^2} \right)
\left(
\frac{i\alpha\epsilon_{\alpha\sigma\beta}p_\sigma}{\sqrt{p^2}} \right)
\left( \frac{i\epsilon_{\beta\kappa\nu} p_\kappa}{p^2} \right)
= \frac{i\alpha\epsilon_{\mu\lambda\nu}p_\lambda}
{\left(p^2\right)^{3/2}} 
\, .
\end{equation}
Since we must have
\begin{equation}
\label{e26}
\int d^3p \frac{p^\lambda e^{i\overline{p} \cdot \overline{r} }}
{p^3} = f \frac{r^\lambda}{r^2} \, ,
\end{equation}
we must have, upon contracting (\ref{e26}) with $r^\lambda$,
\begin{eqnarray}
f & = & -4\pi i \int^\infty_0 dx 
\left( \frac{x\cos x - \sin x}{x^2} \right)
\nonumber \\
& = & 4 \pi i \label{e27} \, .
\end{eqnarray}
Consequently, by (\ref{e25}) and (\ref{e27}), we see that the
one-loop contribution to the $\alpha^1$ term in (\ref{e20}) involves 
having to determine
\begin{mathletters}
\begin{eqnarray}
I^{(1)} & = & -4\pi\alpha \oint_{C} dx^\lambda \oint_{C_f} dy^\sigma 
\frac{\epsilon_{\lambda\kappa\sigma} 
\left( x - y \right)_\kappa}
{| \overline{x} - \overline{y} |^2 } \nonumber \\
& = & -4\pi\alpha \oint_C dx^\lambda \oint_{C_f} dy^\sigma
\epsilon_{\lambda\kappa\sigma} \frac{\partial}{\partial x^\kappa} 
\ln | \overline{x} - \overline{y} | \, , \label{e28} \label{e47a} \\
& = & -4 \pi \alpha \int\!\int_S dS^\lambda \oint_{C_f} dy^\sigma
\left( \frac{\partial}{\partial x^\lambda} \frac{\partial}{\partial x^\kappa}
- \delta_{\lambda\kappa} \frac{\partial^2}{\partial x^2} \right) 
\ln | \overline{x} - \overline{y}| \, , \label{e47b}
\end{eqnarray}
where $S$ is the surface enclosed by $C$.  The first integral in (\ref{e47b})
vanishes as it involves the integral of a total derivative around the closed
curve $C_f$, the second disappears if $dS^\lambda$ is normal to $dy_\sigma$.

We note that, upon applying Stokes' theorem to each of the line 
integrals in (\ref{e47a}), our equation becomes
\begin{equation}
\label{e47c}
I^{(1)} = -8 \pi \alpha \epsilon_{\lambda\kappa\sigma} \int\!\int_S 
dS_\lambda^x 
\int\!\int_{S_f} dS^y_\kappa \, 
\frac{(x-y)_\sigma}{|\overline{x} - \overline{y} |^4}
\, ,
\end{equation}
where $S$ and $S_f$ are two dimensional surfaces bounded by $C$ and
$C_f$ respectively.
\end{mathletters}%

For the  
circle depicted in figure~\ref{exampleframe},
and its framing, the integral of (\ref{e28}) reduces to 
\begin{mathletters}
\begin{eqnarray}
I^{(1)} & = & 4 \pi \alpha 
R\delta \int^{2\pi}_0 d\theta \int^{2\pi}_{0} d\phi \, \frac
{R(1-\cos\theta) \cos\phi - \delta \cos\theta}
{2R^2 + \delta^2 - 2R^2\cos\theta + 2\delta R\cos\phi 
(1-\cos\theta)} \, ,
\end{eqnarray}
which, 
evaluating the integral over $\theta$, becomes
\begin{equation}
= 4 \pi \alpha R \delta \left\{ \frac{2 \pi^2}{\delta} 
\left(1- \sqrt{1-\frac{\delta^2}{R^2}} \right)
- \frac{\pi}{R} \int^{2\pi}_0 d\phi \frac{2R^2 + 3R\delta \cos \phi + \delta^2}
{(R + \delta \cos \phi) \sqrt{4R^2 + 4\delta R \cos \phi + \delta^2} } \right\}
\, ,
\end{equation}
\end{mathletters}%
provided $\delta \neq 0$.  We see that this vanishes as $\delta \rightarrow 0$,
showing that this contribution to the 
dependence on $\alpha$ disappears as the width of the frame shrinks
to zero.

The implementation of the framing 
procedure varies among the various authors.
The method described above is reasonable when one is concerned with
only the $n^{\text{th}}$ term in the expansion 
(\ref{Wilsonloopexpansion}) since
in that case one considers the framing to be 
composed of $n$ ``parallel''
curves; however it is not applicable 
when considering the expansion as 
a whole.   
Hereafter we will adopt a more general interpretation
of simply distributing the 
``topological current'' along the cross section
of an appropriate set of ribbons; 
then only at the end of the calculation
should we 
reduce the ribbons to the original knots through the appropriate
limiting procedure. 
We parameterize the 
frame of a knot $C$ by
\begin{equation}
\label{frame}
\tilde C: \; x^\mu = x^\mu_C(\sigma,t) \;\; \text{ where }  
\, \sigma \in [0,1] 
\, , \;\; t \in [0,1] \, ,
\end{equation}
such that 
$x^\mu_C(\sigma,0) = x^\mu_C(\sigma,1)$, and $x^\mu_C(\sigma,t)$ is
an orientable surface (thereby excluding such ribbons as the M\"obius strip,
which has but one edge.)
Here, we have in mind that $t$ 
should parameterize the length of the loop,
while $\sigma$ corresponds to integration
along the cross-section of the ribbon. 
In order to make the framing of the knot (or, in general the link) 
topologically definite we should also impose the
non-intersection requirement that 
$x_C^\mu(\sigma,t) = x^\mu_{C'}(\sigma',t')$ 
only if $\sigma = \sigma'$, 
$t = t'$ (modulus 1) and $C=C'$. 
Then when we come to regulate the topological
invariants of the theory, we should replace (\ref{Wilsonloop}) by
the following {\em framed} Wilson loop: 
\begin{mathletters}
\label{framedWilsonloop}
\begin{eqnarray}
W[\tilde{C}_R] & = & \tr_R {\cal P} \exp 
\left\{ ig \int_0^1 d\sigma \int_0^1 dt \, \dot{x}_C^\mu(\sigma,t) 
V_\mu(x_C(\sigma,t)) 
\right\} \\
\label{framedWilsonloopexpansion}
& \equiv & \tr_R \sum_{n=0}^\infty (ig)^n 
\int^1_0 d\sigma_n \int^1_0 dt_n \, \dot{x}_C(\sigma_n,t_n) \cdot 
 V(x_C(\sigma_n,t_n)) \nonumber \\
& & \hspace{1cm} \times \int^1_0 d\sigma_{n-1} 
\int^{t_n}_0 dt_{n-1} \, 
 \dot{x}_C(\sigma_{n-1},t_{n-1}) \cdot 
 V(x_C(\sigma_{n-1},t_{n-1}) \\
& & \hspace{2cm} \vdots \nonumber \\
& & \hspace{1cm} \times  \int^1_0 d\sigma_1 \int^{t_2}_0 dt_1 \, 
 \dot{x}_C(\sigma_1,t_1) \cdot V(x_C(\sigma_1,t_1)) 
\,\,\, . \nonumber
\end{eqnarray}
\end{mathletters}%
We will eventually be interested in reducing
the ``width'' of the above framing to zero in order to coincide with 
the original knot. The appropriate limiting
procedure in this case is to replace $\tilde C_R$ 
by $\tilde C_R^\epsilon: x^\mu = x^\mu(\epsilon\sigma,t)$, 
and then take the simple limit $\epsilon \rightarrow 0$ which
has the property that 
$\tilde C_R^\epsilon \rightarrow C_R$.

Now the main point of this section 
concerns the effects that the framing
procedure has on the gauge invariance of the theory.  
The Wilson loop
operator and the pure Chern-Simons action are both
well known to be invariant under 
the infinitesimal gauge transformation 
\begin{equation}
\label{gaugetrans}
V_\mu(x) \rightarrow V_\mu(x) - \case{1}{g} \partial_\mu \omega(x) 
+i [V_\mu(x),\omega(x)] \, ,
\end{equation}
however it can be proven that the framing procedure of 
equation~(\ref{framedWilsonloop}) is {\em not}.  In fact, 
the framed loop operator 
transforms according to
$W[\tilde C_R] \rightarrow W[\tilde C_R] 
+ \delta W[\tilde C_R]$ where
\begin{eqnarray}
\label{framedtrans}
\delta W[\tilde C_R] & = &  
- g \int^1_0 d\sigma \int^1_0 d\sigma' \int^1_0 dt \, 
\dot{x}^\mu(\sigma,t) \tr_R  
W^{x_C(t)}_{x_C(t)}[\tilde C_R] 
\\ & & \hspace{0.5cm} \times
\Big[ V_\mu(x_C(\sigma,t)),
\omega(x_C(\sigma,t))-\omega(x_C(\sigma',t))\Big] \, . \nonumber  
\end{eqnarray}
Here   
$W^{x_C(t)}_{x_C(t)}[\tilde C_R]$ is the 
$R$ representation of the framed
{\em Wilson line} parallel transport operator 
${\cal P} \exp \left[ ig\int^1_0 d\sigma \int^1_0 dt \dot{x}^\mu_C(\sigma,t)
V_\mu(x_C(\sigma,t)) \right]$
along the closed curve $C$ terminating at the point $x_C(t)$. 
It is clear that $\lim_{\epsilon \rightarrow 0} 
\delta W[\tilde C_R^\epsilon]$ vanishes at the classical level.  
At the quantum
level, equation~(\ref{framedtrans}), along with the Nielsen 
transformation
of equation~(\ref{variations}) (in the absence of a background field) 
can be combined to give the following
expression for the gauge 
parameter dependence of the framed Wilson loop:
\begin{eqnarray}
\label{gaugedependence}
\frac{\partial}{\partial\alpha} \langle W[\tilde C_R] \rangle & = &
g^2 \int^1_0 d\sigma \int^1_0 d\sigma' \int^1_0 dt \, 
\dot{x}^\mu(\sigma,t)
\\
& & \hspace{0.5cm} \times \frac{\partial}{\partial\chi} \left\langle 
\tr_R W^{x_C(t)}_{x_C(t)}[\tilde C_R] 
\Big[ V_\mu(x_C(\sigma,t)),
c(x_C(\sigma,t))-c(x_C(\sigma',t))\Big] \right\rangle  \, . \nonumber
\end{eqnarray}
In this latter equation $c(x)$ is 
the ghost field of the BRS multiplet
and the vacuum expectation value
$\frac{\partial}{\partial\chi} \langle \cdots \rangle$ refers to 
the matrix element in the presence of a single operator insertion
$\frac{1}{2\alpha} \overline c \partial \cdot V$ 
(analogous to equation~(\ref{chi})).  
Equation~(\ref{gaugedependence}) follows from taking 
$\frac{\partial}{\partial \chi}$ of (\ref{invariantgamma})
with vanishing background field provided that we identify $\Gamma$
with the vacuum expectation value of the framed Wilson loop 
of~(\ref{framedWilsonloop}).
Whether $\frac{\partial}{\partial\alpha} 
\langle W[\tilde C^\epsilon_R ] \rangle$ generally vanishes in
the $\epsilon \rightarrow 0$ 
limit is unclear to us, but appears to be 
worth further consideration.

\section{Discussion}

In this paper we have calculated part of the two-point function in 
pure non-Abelian
Chern-Simons
theory.   The remaining parts of the two point function have been
computed before, namely the $\alpha$-independent contribution from
the ghost Lagrangian to the modulus 
$\mid \Gamma^{(1)} \mid$ \cite{r14} and
the contribution from the phase of $\Gamma^{(1)}$ \cite{r11}.

The result which we have obtained clearly indicates that this two
point function is {\it both} gauge {\it and} metric
dependent.  This is in contrast to the result in
\cite{Witten} where topological arguments were used to argue
that radiative effects serve only to shift the coupling constant.
The metric dependence of the two point function was noted
previously \cite{r14} where the computation was carried out in the
Landau-Honerkamp gauge (i.e. $\alpha$ = 0) using operator
regularization.  The result of this paper is, to our knowledge, the
first indication of gauge dependence 
in this two point function. 
The regularization procedure cannot be held responsible for this
dependence as it has not been necessary to explicitly introduce any
regulating parameter into the Lagrangian.  Indeed, the dependence
on the gauge parameter has been shown to be consistent with the Nielsen
identities in section~\ref{Nielsenidentities}, above.

The relationship between the vacuum expectation value of Wilson line 
operators and the N-point
correlation functions has been explored in \cite{Guadagnini}.  
The discussion there 
is simplified as these
authors, by using a regulating technique which does not respect BRS 
symmetry in the tree
level
effective action \cite{r6}, 
find that there are no radiative corrections to 
the two and three point
functions
up to two-loop order (and hence that radiative corrections do not 
shift the coupling constant, as
anticipated in \cite{Witten}).  
However, the effect of including the radiative 
corrections found in \cite{r11,r14} and in 
equation~(\ref{e16}) above on the 
vacuum expectations value of framed Wilson line operators 
has, in the preceding section, been shown to give a possible
contribution of order $\alpha$.  This is clearly a problem worth
pursuing.

We note that operator regularization, which is very similar to the 
techniques employed above,
has
been used to demonstrate the absence of any need to renormalize 
Chern-Simons theory at
two-loop order when $\alpha
= 0$ \cite{r18}.

\section{Acknowledgements}

L.Martin and D.G.C.McKeon would like to thank University College 
Galway for its hospitality
while much of this
work was being done.  NSERC provided financial support.  We would 
particularly like to thank
T. Steele for discussions which initiated this investigation.
R. and D. MacKenzie had some interesting comments.


\end{document}